\documentclass[journal]{IEEEtran}
\ifCLASSINFOpdf
\else
\fi
\usepackage{soul}
\usepackage{authblk}
\usepackage{fancyhdr}
\usepackage{bbding}
\usepackage[utf8]{inputenc} 
\usepackage[T1]{fontenc}
\usepackage{multicol} 
\usepackage{algorithm}
\usepackage{algpseudocode}
\usepackage[T1]{fontenc}
\usepackage{graphicx}
\usepackage{xcolor}
\usepackage{wrapfig}
\usepackage{subcaption}

\renewenvironment{IEEEbiography}[1]
  {\IEEEbiographynophoto{#1}}
  {\endIEEEbiographynophoto}

\begin{document}
\chead{This work has been submitted to the IEEE for possible publication. Copyright may be transferred without notice, after which this version may no longer be accessible.
}

%
\title{How to Design Autonomous Service Level Agreements for 6G}

%
%
%

%
\author{\IEEEauthorblockN{Tooba Faisal}
\IEEEauthorblockA{\textit{King's College London, UK, }}
\and
\IEEEauthorblockN{Jose Antonio Ordo\'{n}ez Lucena}
\IEEEauthorblockA{\textit{Telefónica, Spain, }}
\and
\IEEEauthorblockN{ Diego R. Lopez}
\IEEEauthorblockA{\textit{Telefónica, Spain, }}
\and
\IEEEauthorblockN{Chonggang Wang}
\IEEEauthorblockA{\textit{InterDigital, Inc., USA, }}
\and
\IEEEauthorblockN{Mischa Dohler} 
\IEEEauthorblockA{\textit{Ericsson Inc., USA}\thanks{Mischa Dohler contributed as a professor at the King's College London}}}

%

%



\maketitle
\thispagestyle{fancy}
\begin{abstract}
With the growing demand for network connectivity and diversity of network applications, one primary challenge that network service providers are facing is managing the commitments for Service Level Agreements~(SLAs). Service providers typically monitor SLAs for management tasks such as improving their service quality, customer billing and future network planning. Network service customers, on their side, monitor SLAs to optimize network usage and apply, when required, penalties related to service failures. In future 6G networks, critical network applications such as remote surgery and connected vehicles will require these SLAs to be more dynamic, flexible, and automated to match their diverse requirements on network services. Moreover, these SLAs should be transparent to all stakeholders to address the trustworthiness on  network services and  service providers required by critical applications. Currently, there is no standardised method to immutably record and audit SLAs, leading to challenges in aspects such as SLA enforcement and accountability -- traits essential for future network applications. This work explores new requirements for future service contracts, that is, on the evolution of SLAs.
Based on those new requirements, we propose an end to end layered SLA architecture leveraging Distributed Ledger Technology (DLT) and smart contracts. Our architecture is inheritable by an existing telco-application layered architectural frameworks to support future SLAs. We also discuss some limitations of DLT and smart contracts and provide several directions of future studies.
\end{abstract}


%
\IEEEpeerreviewmaketitle


\section{Introduction}
A Service Level Agreement~(SLA) is a contract or agreement between a service provider and a customer. The customer can be an organisation or an individual. It is a legal and detailed document that typically includes roles and responsibilities of all parties involved such as  service quality, duration, and penalties~\cite{jin2002analysis}. Typically, network service providers (e.g., mobile service provider or infrastructure providers) have set procedures in place to monitor their SLAs and generate periodic reports. Such mechanisms allows them to monitor their service quality and keep records in the situations of customer complaints or disputes.

In the 6G networks, many of the existing 5G applications will evolve to be applicable in much more demanding environments. Applications such as remote surgery and extreme uses of the Industry 4.0 paradigms will become  prevalent. Robots and machines will be in charge of critical procedures such as direct health and personal care, and piloting of autonomous vehicles. These procedures will be either human or machine-guided and, either way, they require extremely reliable network connections. That is, future network application providers and users (e.g., surgeons or a car manufacturer) will need to prove the reliability of their applications to gain their customer trust in the product or service. For example, in human-guided remote surgery, the patient and the surgeon will want to ensure that the connection between them is reliable. 

Indeed, this calls for a reliable contractual mechanism for future networks of 6G.  We note two key requirements must be fulfilled for future SLAs. Firstly, the SLAs must be \emph{reliable and trustable}. This means that all the stakeholders should trust the contractual mechanism, and that the SLAs shall be honored in all circumstances. Secondly, they must be \emph{automatically managed} throughout their lifetime, with a special focus on SLA monitoring and enforcement stages. The reason is that future networks will face a high demand for connectivity from an ever-increasing number of devices subscribed to B2B and B2C services. This imposes great scalability burdens that can only be alleviated with automation (zero-touch) procedures, at both network and infrastructure levels.

To that end, we propose the use of DLT, particularly Permissioned Distributed Ledgers (PDLs), to design a contractual mechanism for SLAs in future networks. Distributed Ledgers are an immutable and transparent network of nodes, in which all the participants keep an identical copy of the record.  The planned executions are recorded in the ledgers in the form of executable software code called ``Smart Contracts''. Typically, smart contracts are installed on distributed ledgers and consequently inherit some of the distributed ledgers' properties such as immutability and automated execution. Some properties of smart contracts are briefly described below:


\emph{\textbf{Immutable}} -- Smart contracts are immutable, when they are installed on a ledger, they cannot be amended or deleted.

\emph{\textbf{Auto-executable}} -- When a smart contract is installed on a ledger it becomes auto-executable. 

\emph{\textbf{Transparent}} -- Typically, a smart contract is installed on all  nodes of a distributed ledger system. Therefore, a smart contract becomes visible to all the nodes. 

\begin{figure}
\centering
\begin{subfigure}{0.45\textwidth}
  \includegraphics[width = 80mm]{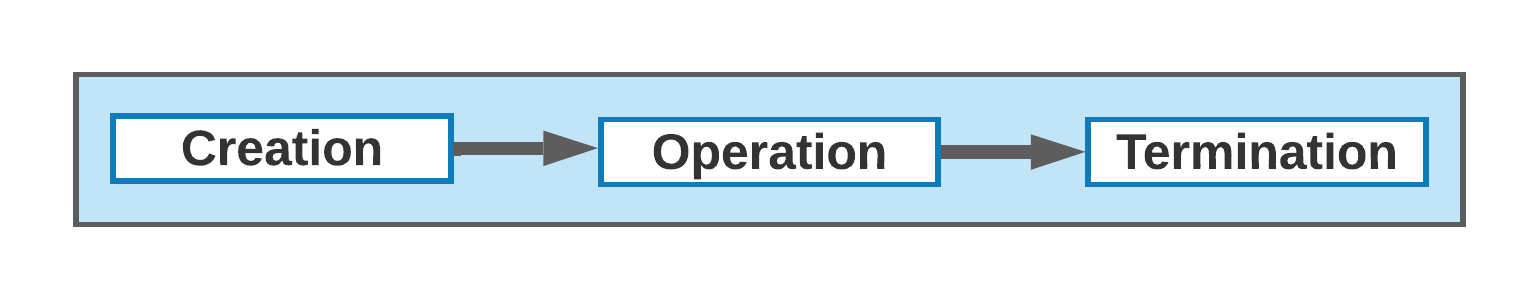}
  \caption{SLA Processes -- Reproduced from ~\cite{wu2012service}}
  \label{fig:(a)sla}
\end{subfigure}

\begin{subfigure}{0.45\textwidth}
  \includegraphics[width = 80mm]{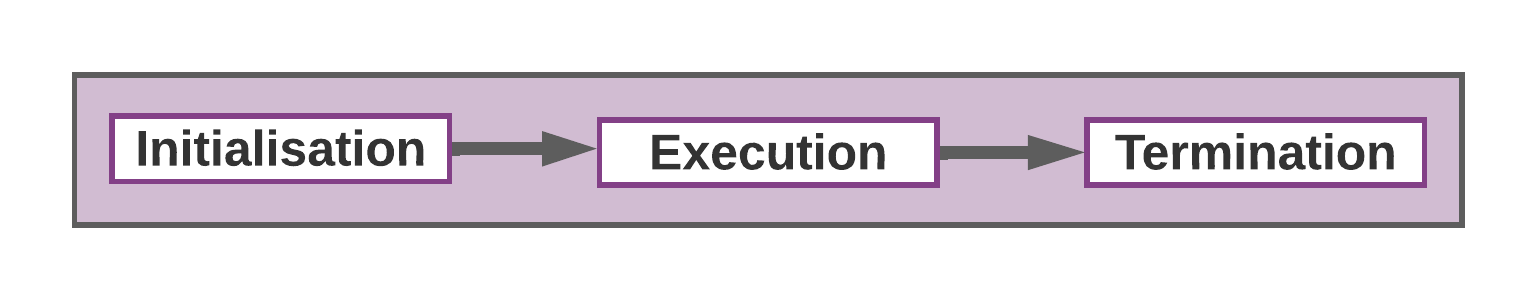}
  \caption{Smart Contract Processes~\cite{etsi_pdl}}
  \label{fig:(b)sc}
\end{subfigure}
\caption{Smart Contract and Service Level Agreements }

\label{fig:sc_processes}
\end{figure}

From lifecycle perspective, smart contracts inherently resembles SLAs as illustrated in Figure~\ref{fig:sc_processes}. For example, an SLA processing lifecycle usually includes a creation phase, an operation phase, and a termination phase; similarly, the life-cycle of a smart contract consists of equivalent phases such as initialisation/creation, execution/logic, and termination. We believe the smart contract lifecycle and its inherent properties (i.e., immutable, auto-executable, and transparent) are aligned with the design requirements of future SLAs. This paper presents a new SLA approach, which leverages smart contracts and PDL to enable autonomous and accountable SLAs for 6G networks. Our contribution primarily includes:
\begin{enumerate}
    \item Novel SLA requirements for future network SLAs are explored and identified. 
    

    \item This new SLA architectural framework is proposed, which utilizes smart contracts to automate the entire multi-layer SLA process.
    \item Several future directions are identified and detailed.
\end{enumerate}

The rest of this paper is organised as follows: Firstly, we review the existing work in Section~\ref{sec:related_work}, Next, we highlight the recommendations about SLA for 6G in Section~\ref{sec:sla_beyond}. We propose a novel DLT-enabled SLA architecture in Section~\ref{sec:arch}. Designing a DLT-enabled architecture is not trivial, several considerations, precautions and planning are required. We discuss these considerations for our architecture in Section~\ref{sec:considerations}. Then, a few future directions on using smart contracts for enabling autonomous SLA for 6G  networks are discussed in Section~\ref{sec:future_work}. Lastly,  we conclude our work in Section~\ref{sec:conclusion}.

%
%
%
%



\section{Related Work}
\label{sec:related_work}
Smart contracts as SLAs have been in discussion for a while, and several applications have been proposed. 
For instance, in \cite{tan2021novel} authors propose the use of DLT in the design of an SLA management system for cloud-based services; however, it is limited to smart contract modularization for SLA in the cloud scenario. Also, they do not discuss the pitfalls of the distributed ledgers and the considerations for their adoption in the architecture. The potential of SLA management in 6G through blockchain/smart contracts is proposed in \cite{hewa2020role} and a platform to manage the SLA between the service providers and customers is presented by~\cite{maksymyuk2021blockchain}. However, none of these define an SLA framework itself with the requirements, challenges, and future directions of SLAs when they are implemented based on smart contracts. A smart contract together with an SLA model is represented by~\cite{zhou2018trustworthy} where the  authors propose a witness committee based model to monitor the SLA; the system incentivizes committee members for being honest. However, they neither discuss the structure of SLA nor challenges related to adopting a smart contract as SLA. Authors in \cite{uriarte2021distributed} propose a model for translating SLAs into a smart contract; the proposed model relies on a third party to collect the monitoring data and adjust the metrics (e.g., billing and payment) accordingly. Like other existing works, the work from~\cite{uriarte2021distributed} is also focused on smart contract implementation for SLAs, but neither discusses the limitations of smart contracts nor the requirements for future networks.

\section{SLAs in  6G}
\label{sec:sla_beyond}
6G networks will introduce a new wave of devices, applications and use cases. 
In this section, we identify the key requirements for future SLAs 

\textbf{Monitorable} -- The future SLAs must be monitorable during their lifetime, that is, from its initialisation to completion; see Figure~\ref{fig:(a)sla}. In future networks, an application such as Industry 4.0 will rely on a an IoT fabric for production and manufacturing services. The most significant challenge with industrial IoT is the management of millions of sensors and monitoring their health. 




\textbf{Transparent} -- A future network SLA will need to be transparently available to all the contract stakeholders for the reasons such as a prospective audit and external SLA verifiability. For example, in a shared network infrastructure like~\cite{faisal2021beat}, all the stakeholders (e.g., service providers and vendors) should have an identical copy of the contract, in which any changes in one contract automatically echoed into other copies of the contract.

\textbf{Auto-Executable} -- Future network SLAs will need to be auto-executable. That is, they must be executable with the pre-programmed conditions without human intervention. We argue that the automated execution should not be from initialisation to termination but maintain certain checkpoints to ensure that micro targets are being met. This means that the management software should verify the contract details at every milestone and take appropriate actions autonomously if any contract violation is identified.

\textbf{Dynamic}  -- SLAs will need  to be dynamically generated and in real-time as per user demand. For example, once a customer orders an autonomous car, the system should generate a contract per his specifications. Special requests such as a change of delivery date and customisation should adjust the main agreed SLA and all other dependent SLAs accordingly and autonomously.

\textbf{Zero Touch Network and Service Management} --
SLAs should be actively monitoring the system and report any anomalies in the system autonomously. In case of potential SLA violations, they must make corrective measurements and adjustments, for instance, to reroute the traffic to another path, and to report this to the control system or the service provider.

\textbf{AI-Enabled} --
\label{subsubsec:ai-enabled}
Managing an ever-growing number of customers will be a significant challenge in the future. The service providers will need to work diligently to resolve customer problems along with the everyday activities such as service requests and QoS monitoring. Indeed, software automation is already a common practice for tasks such as network management. Yet, disputes and disagreements are often done manually and on a case-by-case basis. 


In future generation of contractual systems must have the cognitive ability to make human-like decisions in unforeseen situations like network anomalies and participants' disputes. 

\textbf{Intent-Enabled} -- 
Intent-based management is a potential and prominent shift from complicated service provisioning commands to understandable and straightforward statements referred to as \emph{intents}. As per IETF Definition:  \emph{``An intent specifies the goals and outcomes of the network without specifying how to achieve them''}.

Intent-based management allows the provider to fulfill customer expectation (target service requirements) in a non-prespective manner. In fact, the customer can specify from a wide variety of intents, and the provider has the liberty to allocate resources to achieve those functionalities as per the network feasibility and resource availability. 
\textbf{Cross-Provider Collaboration} -- 
Cross-providers collaboration is already a common practice (e.g., RAN sharing, roaming) to allow operators to extend their service footprint at reasonable costs. However, service providers must adopt this (i.e., collaboration)  practice at a broader scale; instead of installing new infrastructure, first consider collaboration and share the existing front-haul and mid-haul resources mutually. 




\section{Architecture}
\label{sec:arch}

\begin{figure}
    \centering
    \includegraphics[width=80mm]{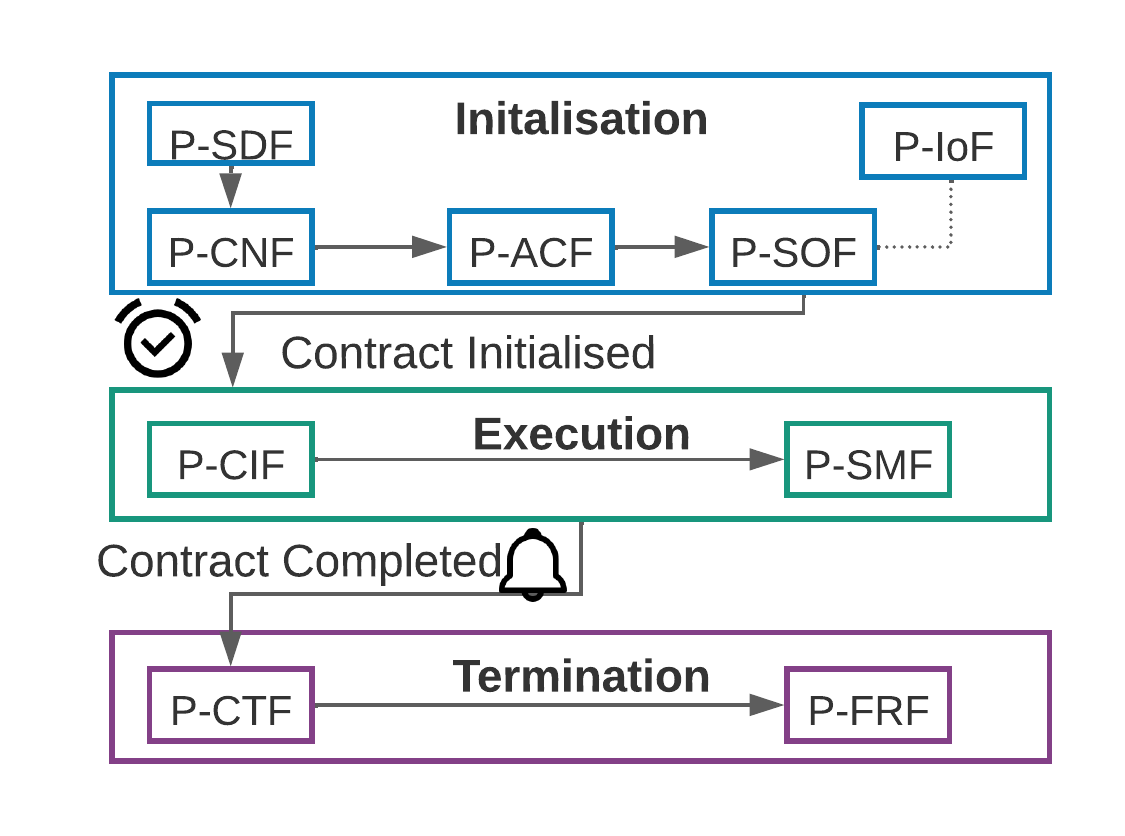}
    \caption{Overall Architecture}
    \label{fig:sla_overall_processes}
\end{figure}

Based on the requirements identified in Section~\ref{sec:sla_beyond}, in this section, we present our  contractual system for future networks of 6G. Our architecture exploits permissioned DLT (i.e., PDL) to enable an automated and transparent contractual tool. As a use case, we position the modularized SLA components in a typical service providers' network slicing architecture~\cite{ordonez2021rollout}.  The components (see Figure~\ref{fig:sla_overall_processes}) are placed in a complete telco stack, that is, in the BSS (Business Support System), OSS (Operation Support System) and network layers~(Figure~\ref{fig:sla_arch}). 
Following are the key players  involved in the architecture:

\textbf{Service Customer} -- including end-users (B2C market) and most notably enterprise customers (B2B/B2B2C market), such as hyperscalers and  large-scale content service providers.

\textbf{Governance} -- A group of players, which oversees the network management and operations; they also maintain the compliance strategies within the architecture and watches for wrongdoings. There are two types of governance involved in the architecture: 1) \textbf{Local Governance} that belongs to one network architecture only; and 2) \textbf{Inter-PDL Governance} which sits atop more network architectures and performs the management tasks.

Typically, an SLA has several modules or subcontracts, which form an end-to-end SLA or contract.  The most common subcontracts in our architecture are detailed below. Note that, in this work, we provide a generic contractual system architecture. As per the application, a contractual mechanism can have more/fewer subcontracts.

\textbf{Intent-Translation Smart Contract (IT-SC)} -- records the agreed SLA terms established on the intent to the ledger. The IT-SC is executed by the ITF (Intent Translation Function)  
    
\textbf{Access Control Smart Contract (AC-SC)} -- records the access control credentials to the ledger. It includes an internal timers, which automatically revokes the access rights when the assigned access time elapses. The AC-SC is executed by the P-ACF (PDL Access Control Function) and 
    
\textbf{Service Orchestration Smart Contract (SO-SC)} -- sets up the service provisioning through resource allocation; for instance, types and duration of resources used. It is executed by the P-SOF (PDL Service Orchestration Function).  

\textbf{Service Recording Smart Contract (SR-SC)} -- records the service  usage to the ledger. It is executed by executed by the PDLF (PDL Function).

Our architecture is underpinned by the following strata, which further includes functional elements of the architecture:
\subsection{Orchestration Stratum}
It is the first point of contact from a user to the network and part of the BSS and OSS Layers. A customer requests for the services (e.g., a network slice or RAN) through the orchestration stratum. It has the following functions:


\subsubsection{Intent Translation Functions (ITF)}
The first phase in our architecture is to interpret the customers' intent. Note that, in the future networks, the network service requests do not need to be described in or include network-specific vocabulary (e.g., bandwidth and packet loss rate), and the customers will have the liberty to express their requirements in simple terms (i.e., \emph{Intent}) (Section~\ref{sec:sla_beyond}).

To this end, ITF translates intents to the actionable or configuration commands. Intent translation is usually an iterative process, and the ITF may need to consult the customer/user several times before the terms are agreed upon. The final agreement on the intent translation is recorded through a smart contract to the ledger. 


\subsubsection{PDL Service Discovery Function~(P-SDF)}
Once the customer agrees on the translated intent values, the P-SDF will browse through the SLA catalogue and show the available services to the customer as per the stated requirements. For example, if a customer has asked for a network slice for a football match, the P-SDF will show all the available service offerings suitable for the football match during the time and the day. 

\subsubsection{PDL Contract Negotiation Function (P-CNF)}
In some cases, it may be required to negotiate the SLAs, for example, if the service provider wishes to provision their services through an auction or when a customer needs unique services. P-CNF will provide negotiation functionalities such as bids and customised pricing in such a situation. The governance of the PDL controls this function, but the service provider will negotiate the SLA. 
\subsubsection{PDL Access Control  Function (P-ACF)}
In the next step, internal PDL governance will assign access credentials to the customers through P-ACF and as per the agreed SLA. The access control will be recorded to the ledger through AC-SC for a limited and SLA-agreed time. 

\subsubsection{PDL Service Orchestration Function (P-SOF)}

PDL Service Orchestration Function allocates the resources as per the agreed SLA. For example, a shared Radio Unit. Note that services are orchestrated at this point only and provisioned after the contract initialisation.

\subsubsection{PDL Contract Initialisation Function~(P-CIF)}
Once the services are orchestrated, the P-CIF initialises the smart contract (i.e., SLA), and the services are provided. P-CIF also initialises a timer within the contract to keep track of service provisioning duration. 


\subsubsection{PDL Inter-Operability Function (P-IoF)}
We advocated a single multi-operator and shared architecture and argued on the viability of service providers' collaboration~(Section~\ref{sec:sla_beyond}). However, several networks are likely to be involved in some service provisioning, for example, in a roaming scenario. Yet, distributed ledgers have a significant consideration of interoperability. That is, not every DLT protocol is compatible with other DLT protocols. To this end, P-IoF will ensure that appropriate functionalities are implemented to enable interoperability between two service providers. The PDL interoperability strategies and algorithms are a research area on their own, and we leave this for future work. 
\subsection{Functional Stratum}
\subsubsection{PDL Function (PDLF)}
\begin{figure*}
    \includegraphics[width=\textwidth]{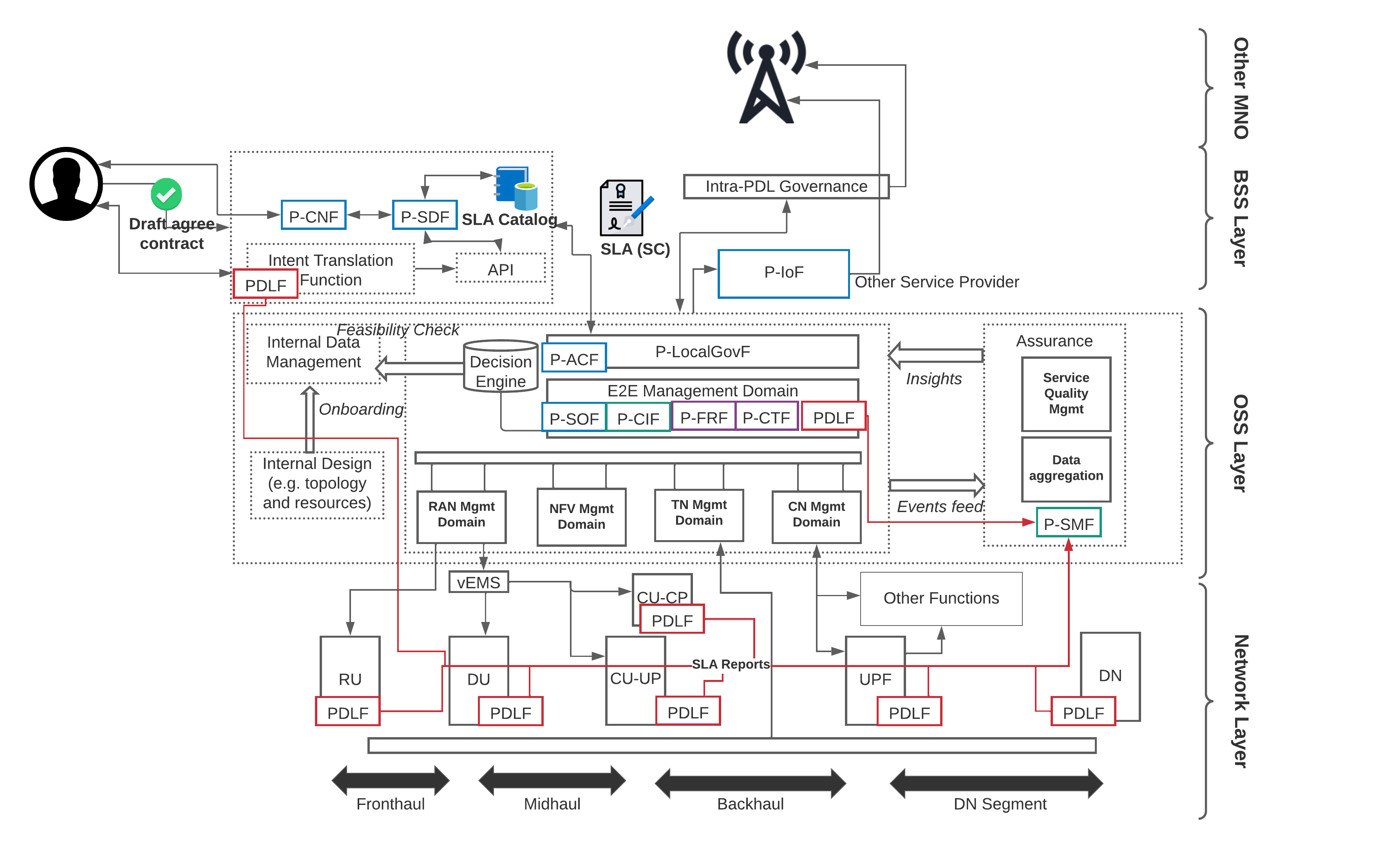}
    \caption{PDL-Enabled SLA Modular Architecture Built atop Network Slicing Use Case~\cite{ordonez2021rollout}}
    \label{fig:sla_arch}
\end{figure*}

This function provides the PDL functionalities, that is, recording the data to the PDL and running the consensus mechanism. Every operator-managed resource, including network functions and OSS assets, is equipped with a PDLF (see Figure~\ref{fig:sla_arch}). However, the PDL functionalities are not always required to be active and are only activated when needed. For instance, if the RAN usage is needed to be stored, the PDLF in the Radio Unit (RU) and Distributed Unit (DU) will be active. The overall idea is that every operator-managed resource is DLT-enabled and ready to execute smart contracts. 

A PDLF will have two major functions 1) record the data to the PDL, and 2) generate and send periodic reports to the intra-PDL governance for future auditability. 


\subsubsection{PDL SLA Management Function~(P-SMF)}
This function monitors the SLA and collects the insights of SLA execution and checks for anomalies. In case any misbehaviour is identified, the P-SMF will immediately send a report to the intra-PDL governance, and the smart contract will be interrupted immediately through control instructions~\cite{etsi_pdl}.

\subsection{Termination Stratum}
Once the service contract is completed, and SLA is finalised, the smart contract must be deactivated properly~\cite{etsi_pdl}. The complete and safe termination of a contract is the responsibility of the Termination Stratum and has the following functions:
\subsubsection{PDL Contract Termination Function (P-CTF)}
 All the smart contract variables are cleared at this stage. The access rights are automatically revoked by the P-ACF, however, the governance will ensure that access rights are revoked and the smart contract is no longer accessible by the any stakeholder.
\subsubsection{PDL Final Report Function (P-FRF)}
 The system will also generate a final report. This report will include SLA lifecycle parameters such as the SLA start and end times, proof of successful deactivation and details of the involved parties. The governance of PDL, will keep these reports for the audit purposes. Note that, the PDL, itself is immutable, therefore some of the details such as execution times and parameters will be recorded in the PDL. However, we propose to generate these reports for two reasons: 1) first, all the details such as confirmation of access rights revocation will not be the part of the PDL; and 2) secondly, in some cases the stakeholders may choose to execute the smart contract with encrypted/hashed parameters. Keeping a record of SLA reports will provide detailed results for the stakeholders. 


\section{Considerations}
\label{sec:considerations}


This section highlights the considerations necessary for adopting smart contracts as SLAs.
\subsection{SLA Enforcement}
Distributed ledgers are a network of world-wide set of distributed nodes. When an SLA is installed on the ledger and the nodes are distributed in a global footprint, different jurisdiction rules apply to the node and smart contract enforcement across the national borders is still a challenge. For example, certain GDPR laws apply to the EU and the UK but may not be applicable to the other countries.
\subsection{SLA Composability}

An SLA has horizontal and vertical components~\cite{1045756}. The horizontal SLAs represents an agreement whereby provider and customer roles correspond to actors from the same layer, e.g., contract between two network service providers.  The vertical SLA represents an agreement whereby provider and customer role corresponds to actors from different layers e.g., the contract between the vendor and service provider. These horizontal and vertical components are chained together as functions to form an end-to-end SLA. During service provisioning, one SLA function would often call other SLA functions.

%


Indeed, smart contracts can automate this end-to-end process. They are auto-executable. This means an uncontrolled or unmanaged function can initiate a chain of authorised and unauthorised executions. We have witnessed, in the famous Decentralised Autonomous Organisation (DAO) attack, developer mistakes that led the attackers to access the payment functions of the SLA and cost huge monetary losses. To enable security in smart contracts, they should be coded so that they are allowed to access other functions with a comprehensive security framework including strict access-control~\cite{etsi_pdl}.
\subsection{AI-Enabled Smart Contracts}
As discussed in Section~\ref{subsubsec:ai-enabled}, future SLAs may need to be AI-enabled. 
Yet, computations required for AI-based systems would require compatible hardware and software. Generally, the distributed ledger nodes' processing resources are dominated by the node management tasks such as transaction processing. 
One of the possible solution is to offload  computational intensive task to an external system~\cite{etsi_pdl}. With modularization, a smart contract can be divided into several small modules, and computation-intensive parts can be installed on appropriate hardware. 
\subsection{Inherent Properties}
\label{subsec:inherent_properties}
\subsubsection{Immutability and Scalability}

Despite its advantages, the challenges with immutability are two-fold. First, the immutability leads to scalability limitation, that is, if a contract cannot be deleted or amended, erroneous and expired contracts will stay in the ledger forever. Secondly, it is also a security risk. For example, a developer codes a smart contract and forgets to make the payment function private (i.e., accessible by the owner only). Malicious users can access the smart contract and drain the contract's funds. Because smart contracts are unamendable, they cannot be updated or \emph{patched} and will stay accessible by every member of the ledger. If a term in an SLA is wrongly coded, it will stay in the ledger forever. The possible solutions to these problems are careful planning and \emph{Termination Function}~\cite{etsi_pdl}. Note that the termination functions cannot remove a smart contract from the ledger but will clear all the variables and deactivate it, therefore scalability challenges remain. Solutions such as off-chain storage~\cite{eberhardt2018off}\cite{etsi_pdl} are proposed by the research community to solve scalability challenges.





\subsubsection{Transparency}

In distributed ledgers, every node keeps a copy of the ledger, and all the transactions and smart contracts are replicated across the network.  This is problematic in the scenario when a number of competitors are in the ledger working as a node. For example, if vendor A has some contract with Operator A' and does not want to disclose their contractual conditions to operator B and vendor B'. Techniques such as Hyperledger Fabric's Channels~\cite{hlf_channels} allows the PDL users to create private channels of communication within a ledger network.
\subsection{Data Inputs}

Data in smart contracts is entered automatically, which means that it is injected through either  another smart contract or an external oracle. The whole process is automated, in the sense that there is no way to pick up if the internal or external sources are entering the wrong data in the smart contracts. 
In oracles, for instance,  data is collected from several sources (e.g., weather feeds) and if those sources are
malicious it will be difficult to identify the problem.

Secure data feed proposals such  as Town Crier~\cite{zhang2016town} can be used but the limitations and requirements of these mechanisms should be considered. Another option is an internal oracle service as proposed in~\cite{etsi_pdl}; whereby the service is managed and maintained by the distributed ledger network and overseen by the governance.
\subsection{Interoperability}
An SLA generally involves two or more organisations.
If two organisations operate different ledger types, the apparent challenge is interoperability between the ledger types. Typically, distributed ledgers vary in several ways, such as consensus algorithms and access control mechanisms, and interoperability between them have challenges such as synchronisation latency and accurate and timely translation of parameters.

Strategies such as  Notary schemes, Relay schemes or Hash-locking~\cite{koens2019assessing} are discussed in the literature and can be adapted to enable interoperability between distributed ledgers. However, network designers should be careful to guarantee that both ledgers are synchronised in a timely manner so that SLA integrity and enforcement is not affected. 
\subsection{Security}
Security is of paramount importance for SLAs; if an SLA is tampered with or executed with inaccurate data, it can cause monetary losses such as payment to illegitimate parties. Particular care must be taken when SLAs are coded as smart contracts because smart contracts are vulnerable to several security problems such as their inherent properties (see Section~\ref{subsec:inherent_properties}) and hardware security challenges. Hardware or physical layer attacks, for instance, Man-in-the-Middle Attack, can be protected by holding the device owner accountable for any wrongdoings. The governance of the ledger can schedule periodic security audits of the nodes to ensure that the device is secure and not being tampered with.


\section{Future Directions \& Challenges}
\label{sec:future_work}
\subsection{Open and Distributed Networks}

Future 6G networks will be more open and agile. This means that i) network functions will be more plug-and-play, thanks to the open interfaces; and ii) players will be able to join the network or leave the network without any cumbersome process. SLA through smart contracts can facilitate automation in the process. For example, a customer joining the network will be registered automatically through the execution of a smart contract. The main challenge would be that devices (acting as nodes in a PDL environment) will be more agile, that they will often be joining the network and leaving the network simultaneously. 

Typically, distributed ledgers require consensus to approve/reject the transaction, and this means that if the consensus is not achieved (e.g., because there are not enough nodes active in the network to validate the transaction), the PDL will not reach to a consensus and transaction will be rejected. The second problem is that if the majority of the nodes collude against the newly joined nodes and reject their transactions, the system will never be impartial; that is, honest nodes will be penalised. 

A potential solution is to design an appropriate consensus mechanism for the PDL and a strong governance model with a governance body~(e.g., Ofcom) overseeing the whole system. The Governance's authority can enforce strategies to ensure a threshold number of nodes stay in the system. They can also take compliance  actions the misbehaving nodes such as blacklisting the colluding nodes.
\subsection{Integration of Non-Public (Private) and Public Network}
Private 5G networks, i.e., non-public networks, are getting attention.  By providing authority over wireless coverage and capacity, the private 5G network market ensures guaranteed and secured connectivity, while supporting a wide range of applications, ranging from push-to- talk group communications and real-time video delivery to wireless control and automation in industrial environments. This has motivated to a wide variety of industry verticals (e.g., militaries, utilities, public safety agencies, manufacturers) to start making sizeable investments in private 5G networks. 

Though early roll-out of private 5G services will be based on Standalone Non-Public Networks (SNPN), the midterm scenarios will build on Public Network Integrated Non-Public Network (PNI-NPNs). In this modality, the private 5G network is provisioned with the support of public network (PLMN) resources. The role of PLMN here is to provide service continuity for those cases where the user moves out of private network coverage.  For example, a hospital has network connectivity through a non-public network, and an automated ambulance is dispatched to collect the patients from an accident site. The ambulance will likely move beyond the coverage of the non-public network and will enter to the public network managed coverage. In such a situation, the QoS will be affected, taking into account that non-public networks typically are designed to provide better network guarantees than public networks. For this hybrid (private-public) model, it is needed to formulate the SLA as a weighted composition of NPN's SLA and PLMN's SLA, with PLMN's SLA set based on negotiation between the mobile network operator and the hospital. 

In future 6G networks, integration and interoperability between private and public networks will happen at a much wider scale, connecting together networks of different nature and scale, thereby realizing the so-called network of networks. In these scenarios, SLAs will be defined as a composition of a number of fine-grained, context-aware SLAs, with much more variants that shall be managed appropriately.  



\section{Conclusion}
\label{sec:conclusion}
This work highlights the requirements of the SLAs for future 6G networks. Based on the identified requirements, we argued that smart contracts are the key to future SLAs. They pose the essential properties for future SLAs, and on this notion,  we proposed a modularised, PDL-enabled SLA architecture for future networks of 6G. Like every technology, distributed ledgers have limitations, which may hinder their adoption as SLAs. We discussed these limitations in detail and argued that these limitations could be managed and mitigated through comprehensive planning and standardization. We believe this work will mark a new era of SLAs which are trustable by all the stakeholders.


%




\ifCLASSOPTIONcaptionsoff
  \newpage
\fi



%
\bibliographystyle{IEEEtran}
\bibliography{bib/ref.bib}

\begin{thebibliography}{10}
\providecommand{\url}[1]{#1}
\csname url@samestyle\endcsname
\providecommand{\newblock}{\relax}
\providecommand{\bibinfo}[2]{#2}
\providecommand{\BIBentrySTDinterwordspacing}{\spaceskip=0pt\relax}
\providecommand{\BIBentryALTinterwordstretchfactor}{4}
\providecommand{\BIBentryALTinterwordspacing}{\spaceskip=\fontdimen2\font plus
\BIBentryALTinterwordstretchfactor\fontdimen3\font minus
  \fontdimen4\font\relax}
\providecommand{\BIBforeignlanguage}[2]{{%
\expandafter\ifx\csname l@#1\endcsname\relax
\typeout{** WARNING: IEEEtran.bst: No hyphenation pattern has been}%
\typeout{** loaded for the language `#1'. Using the pattern for}%
\typeout{** the default language instead.}%
\else
\language=\csname l@#1\endcsname
\fi
#2}}
\providecommand{\BIBdecl}{\relax}
\BIBdecl

\bibitem{jin2002analysis}
L.-j. Jin, V.~Machiraju, and A.~Sahai, ``{Analysis on Service Level Agreement
  of Web Services},'' \emph{HP}, vol.~19, pp. 1--13, 2002.

\bibitem{wu2012service}
L.~Wu and R.~Buyya, ``{Service Level Agreement in Utility Computing Systems},''
  in \emph{Performance and dependability in service computing: Concepts,
  techniques and research directions}.\hskip 1em plus 0.5em minus 0.4em\relax
  IGI Global, 2012.

\bibitem{etsi_pdl}
``{Details of ETSI Work Item: DGS/PDL-0011 Specifications of Requirements of
  Smart Contracts' Architecture and Security},'' \url{https://bit.ly/3qGIZiq},
  accessed: 17-11-2021.

\bibitem{tan2021novel}
W.~Tan, H.~Zhu, J.~Tan, Y.~Zhao, L.~D. Xu, and K.~Guo, ``{A Novel Service Level
  Agreement Model Using Blockchain and Smart Contract for Cloud Manufacturing
  in Industry 4.0},'' \emph{Enterprise Information Systems}, pp. 1--26, 2021.

\bibitem{hewa2020role}
T.~Hewa, G.~G{\"u}r, A.~Kalla, M.~Ylianttila, A.~Bracken, and M.~Liyanage,
  ``{The Role of blockchain in 6G: Challenges, opportunities and research
  directions},'' in \emph{6G SUMMIT}.\hskip 1em plus 0.5em minus 0.4em\relax
  IEEE, 2020, pp. 1--5.

\bibitem{maksymyuk2021blockchain}
T.~Maksymyuk, M.~Volosin, J.~Gazda, and M.~Liyanage, ``{Blockchain-based
  Decentralized Service Provisioning in Local 6G Mobile Networks},'' in
  \emph{ACM SenSys}, 2021, pp. 516--519.

\bibitem{zhou2018trustworthy}
H.~Zhou, C.~de~Laat, and Z.~Zhao, ``Trustworthy cloud service level agreement
  enforcement with blockchain based smart contract,'' in \emph{IEEE CloudCom},
  2018.

\bibitem{uriarte2021distributed}
R.~B. Uriarte, H.~Zhou, K.~Kritikos, Z.~Shi, Z.~Zhao, and R.~De~Nicola,
  ``Distributed service-level agreement management with smart contracts and
  blockchain,'' \emph{Concurrency and Computation: Practice and Experience},
  vol.~33, no.~14, p. e5800, 2021.

\bibitem{faisal2021beat}
T.~Faisal, M.~Dohler, S.~Mangiante, and D.~R. Lopez, ``{BEAT:
  Blockchain-Enabled Accountable and Transparent Network Sharing in 6G},''
  \emph{arXiv preprint arXiv:2107.04328}, 2021.

\bibitem{ordonez2021rollout}
J.~Ordonez-Lucena, P.~Ameigeiras, L.~M. Contreras, J.~Folgueira, and D.~R.
  L{\'o}pez, ``On the rollout of network slicing in carrier networks: A
  technology radar,'' \emph{Sensors}, vol.~21, no.~23, p. 8094, 2021.

\bibitem{1045756}
E.~Marilly, O.~Martinot, S.~Betge-Brezetz, and G.~Delegue, ``{Requirements for
  Service Level Agreement Management},'' in \emph{IEEE IPOM}, 2002, pp. 57--62.

\bibitem{eberhardt2018off}
J.~Eberhardt and J.~Heiss, ``Off-chaining models and approaches to off-chain
  computations,'' in \emph{SERIAL}, 2018, pp. 7--12.

\bibitem{hlf_channels}
{Hyperledger Fabric}, ``{Channels},'' \url{https://bit.ly/3eyaa7X}, accessed:
  22-12-2021.

\bibitem{zhang2016town}
F.~Zhang, E.~Cecchetti, K.~Croman, A.~Juels, and E.~Shi, ``{Town Crier: An
  Authenticated Data Feed for Smart Contracts},'' in \emph{ACM SIGSAC}, 2016,
  pp. 270--282.

\bibitem{koens2019assessing}
T.~Koens and E.~Poll, ``Assessing interoperability solutions for distributed
  ledgers,'' \emph{Pervasive and Mobile Computing}, vol.~59, p. 101079, 2019.

\end{thebibliography}
%
\vspace{-0.3cm}

\begin{IEEEbiography}{Tooba Faisal} is a PhD student at KCL working on Telco-blockchain. Her current research interests are designing secure smart contracts, DLT and 6G networks. She also has a MRes in Security Science from UCL.
She is a KCL's delegate in the ETSI ISG on PDL and rapporteur of smart contract GR and GS. 
\end{IEEEbiography}
\vspace{-0.3cm}

\begin{IEEEbiography} {Jose Ordonez-Lucena} joined Telefónica I+D in 2018 as a Technology Analyst and Standards Specialist, within the Global CTIO Unit. He is currently involved in technology exploration activities for B5G systems, with a special emphasis on end-to-end network slicing solutions and APIs for network capability exposure, considering their joint applicability for public-private network integration scenarios. 
\end{IEEEbiography}
\vspace{-0.3cm}
\begin{IEEEbiography}{Diego R. Lopez}
 joined Telefónica I+D in 2011 as a Senior Technology Expert on network middleware and services. He is currently in charge of the Technology Exploration activities within the GCTO Unit of Telefónica I+D.
 Diego is currently focused on identifying and evaluating new opportunities in technologies applicable to network infrastructures.
 
\end{IEEEbiography}
\vspace{-0.3cm}
\begin{IEEEbiography} {Chonggang Wang}
 is currently a Principal Engineer with InterDigital, Inc. He has more than 20 years of experience in the fields of wireless communications, networking, and computing, including research, development, and standardization. He is the Founding Editor-in-Chief of the IEEE IoT Journal and is currently the Editor-in-Chief of IEEE Network. He is an IEEE Fellow.
\end{IEEEbiography}
\vspace{-0.3cm}
\begin{IEEEbiography}{Mischa Dohler} is the Chief Architect in Ericsson and visiting Professor in wireless communications at KCL, driving cross-disciplinary research and innovation in technology, sciences, and arts. He is a Fellow of the IEEE, the Royal Academy of Engineering, the Royal Society of Arts (RSA) and the Institution of Engineering and Technology (IET).
\end{IEEEbiography}
\end{document}